\documentclass[12pt]{article}
\usepackage{graphicx, amsmath, amssymb, cite, setspace, color,fancyheadings, sidecap}
 \usepackage[top=1 in, bottom=1 in, left=1 in, right=1 in]{geometry}

\newcommand{\captionfonts}{\footnotesize} %make footnote font size smaller
\makeatletter 
\long\def\@makecaption#1#2{%
  \vskip\abovecaptionskip
  \sbox\@tempboxa{{\captionfonts #1: #2}}%
  \ifdim \wd\@tempboxa >\hsize
    {\captionfonts #1: #2\par}
  \else
    \hbox to\hsize{\hfil\box\@tempboxa\hfil}%
  \fi
  \vskip\belowcaptionskip}
\makeatother

\def\fnote#1#2{\begingroup\def\thefootnote{#1}\footnote{#2}
     \addtocounter{footnote}{-1}\endgroup}

\begin{document}
\title{The Case of the Disappearing Instanton}

\author{Adam~R.~Brown$^{1,2}$ \,and Alex~Dahlen$^{1}$ \vspace{.1 in}\\  
\vspace{-.3 em}  $^1$ \textit{\small{Physics Department, Princeton University, Princeton, NJ 08544, USA}}\\
\vspace{-.3 em}  $^2$ \textit{\small{Princeton Center for Theoretical Science, Princeton, NJ 08544, USA}}}
\date{}
\maketitle
\fnote{}{emails: \tt{adambro@princeton.edu, adahlen@princeton.edu}}

\begin{abstract}
\noindent 
Instantons are tunneling solutions that connect two vacua, and under a small change in the potential, instantons sometimes disappear.  
We classify these disappearances as \emph{smooth} (decay rate $\rightarrow 0$ at disappearance) or \emph{abrupt} (decay rate $\neq 0$ at disappearance).  Abrupt disappearances mean that a small change in the parameters can produce a drastic change in the physics, as some states become suddenly unreachable. 
The simplest abrupt disappearances are associated with annihilation by another Euclidean solution with higher action and one more negative mode; higher-order catastrophes can occur in cases of enhanced symmetry.
We study a few simple examples, including the 6D Einstein-Maxwell theory, and give a unified account of instanton disappearances. 
\end{abstract}
\maketitle

%When a quiet corner of the string theory landscape is disturbed by the disappearance of one its favorite instantons, we are on the case.
\vspace{10cm}
\thispagestyle{empty}
\pagebreak

\section{Introduction} \label{sec:introduction}

Instantons are Euclidean solutions that mediate tunneling between two vacua. Under a smooth change in the potential instantons sometimes disappear.  There are two possibilities.

\begin{figure}[h!] %  figure placement: here, top, bottom, or page
   \centering
   \includegraphics[height=2in]{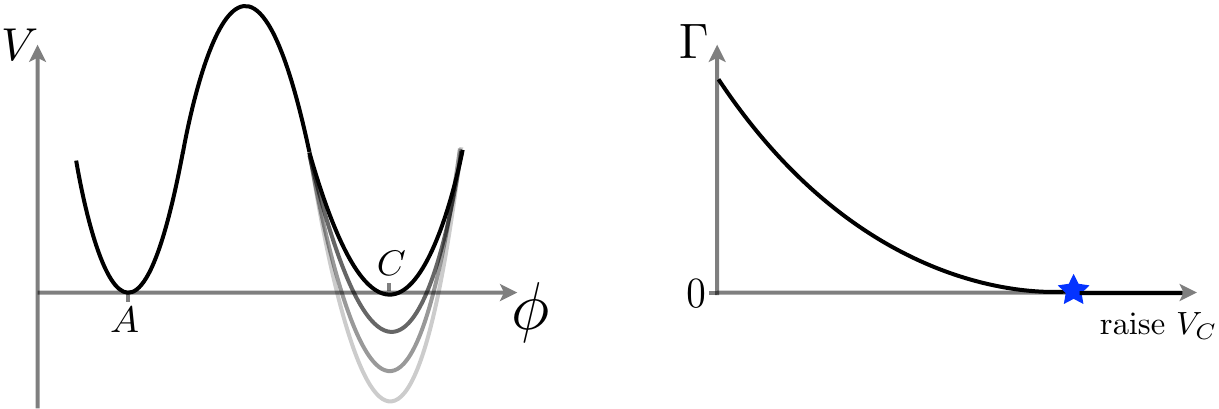} 
   \caption{As $V_C$ is raised, the $A \rightarrow C$ tunneling rate $\Gamma$ goes smoothly to zero as the $AC$ instanton disappears. The disappearance is denoted by a star. In flat spacetime, the instanton disappears when the vacua become degenerate.}
   \label{fig:smoothpotential}
\end{figure}

 \textbf{Smooth disappearance:} As the instanton disappears, the tunneling rate $\Gamma \rightarrow 0$. 
An example of a smooth disappearance is given in Fig.~\ref{fig:smoothpotential}. As $C$ is raised towards $A$, the rate to tunnel from $A$ to $C$ goes smoothly to zero (for field theory in flat spacetime). 

\begin{figure}[h!] %  figure placement: here, top, bottom, or page
   \centering
   \includegraphics[height=2in]{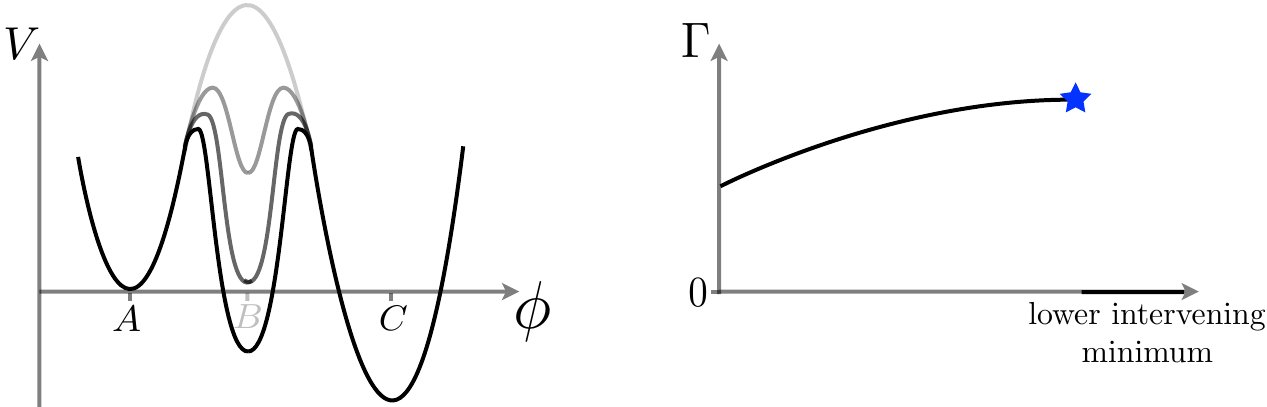} 
   \caption{As a new minimum is lowered, the direct $A \rightarrow C$ tunneling rate $\Gamma$ jumps abruptly to zero as the $AC$ instanton disappears. This happens after $B$ becomes degenerate with $A$, but before $B$ becomes degenerate with $C$.}
   \label{fig:abruptpotential}
\end{figure}

\textbf{Abrupt disappearance:} As the instanton disappears, the tunneling rate $\Gamma > 0$, right up to the moment of disappearance.  An example of an abrupt disappearance is given in Fig.~\ref{fig:abruptpotential}. As we will see in Sec.~\ref{subsec:twostep}, adding an intermediate minimum can make the instanton mediating direct decay from $A$ to $C$ abruptly disappear at nonzero $\Gamma$. 

\pagebreak 
In this example, even after the $A \rightarrow C$ instanton has abruptly disappeared the field can still reach $C$ indirectly, via $A \rightarrow B$ then $B \rightarrow C$. But we will see examples where the abruptly disappearing instanton was the \emph{only} way to reach $C$, meaning the field will now never reach $C$, not even indirectly. A small change in the parameters of the theory has brought about a dramatic change in the physics.

That interpolating instantons do not exist between all pairs of vacua has been noted already in the literature \cite{Cvetic:1994ya,Johnson:2008vn,Aguirre:2009tp,Brown:2010bc} . In this paper, we study the smooth changes in the potential that take these instantons from existing to not existing, that make them disappear. 
Section~\ref{sec:review} gives a brief review of tunneling instantons. Section~\ref{sec:simple} studies two simple field theory examples of abrupt disappearances and finds they exhibit three salient characteristics:
\begin{enumerate}
\item There's always another solution involved. An instanton can only abruptly disappear when annihilated by another Euclidean solution. 
\item This other solution always has higher action and an extra negative mode. 
\item An instanton can only abruptly disappear if it is a subdominant decay mode out of $A$ (though it may be the dominant, or only, decay mode into $C$). 
\end{enumerate}

Section~\ref{sec:summary} gives a unified account of this behavior. We treat smooth and abrupt disappearances together and show how they arise in the saddle-point approximation to the path integral. We also discuss moving away from the semiclassical limit, which softens the abruptness of the disappearance by an amount of order $\hbar$, so that the transition from tunneling to no tunneling is sharp but continuous. Section~\ref{sec:complex} looks at gravitational instantons, which have both smooth and abrupt disappearances. We will also see our first higher-order disappearance: because of its enhanced symmetry, the Hawking-Moss instanton annihilates other solutions in a cusp catastrophe.   Finally, in the Appendix, we consider in detail  the landscape of 6D Einstein-Maxwell theory, which has received much attention recently as a toy string theory landscape \cite{BlancoPillado:2009di}.  This model exhibits a rich structure of disappearances.

The disappearance of instantons has potentially alarming consequences for the ability of eternal inflation to populate the string theory landscape in the early Universe. After all, in flat spacetime `no instanton' can mean `no transition', so that regions of flat space landscapes really can be inaccessible. But we will argue in an upcoming paper \cite{us} that even without an instanton, and even with intervening anti-de Sitter sinks, there is always a nonzero rate, no matter how tiny, for a de Sitter vacuum to transition to any other de Sitter vacuum in the landscape.

\pagebreak
\section{Review  of Tunneling Instantons} 
\label{sec:review}

A field in the $A$ vacuum is classically stable, but quantum mechanically unstable: it may tunnel out by nucleating a bubble of $C$. 
\begin{figure}[h]%  figure placement: here, top, bottom, or page
   \centering
   \includegraphics[height=1.9in]{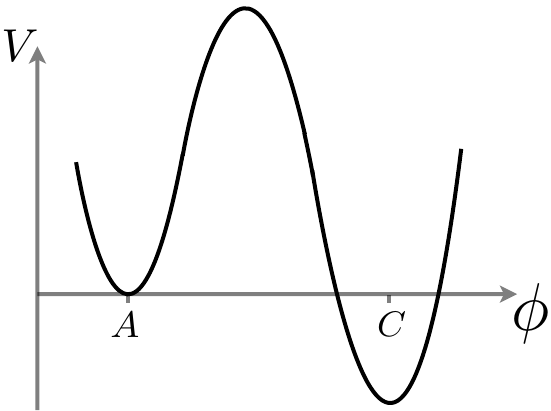} 
   \caption{Tunneling from $A$ to $C$ proceeds by bubble nucleation. In the semiclassical description the field makes a quantum jump: a zero-energy bubble of  true vacuum $C$ appears in the false vacuum $A$. The bubble then classically grows, completing the transition.}
   \label{DIfig:potential}
\end{figure}
In the saddle-point approximation to the path integral, tunneling is governed by an instanton, a classical Euclidean solution that connects the two vacua. To leading semiclassical order in $\hbar$, the corresponding rate per unit volume $\Gamma$ is
\begin{equation}
\Gamma \sim e^{- \Delta S / \hbar} ,
\end{equation}
where $\Delta S$ is the Euclidean action of the instanton minus the Euclidean action of the false vacuum. 

The instanton that mediates this decay has the form of an O(4)-symmetric bubble of true vacuum separated from the false-vacuum exterior by an interpolating wall.  The Euclidean action of a scalar field with this symmetry in flat Euclidean space is
\begin{equation}
S = 2 \pi^2 \int d \rho \, \rho^3 \left( \frac{1}{2} \dot{\phi}(\rho)^2  + V(\phi) \right) ,
\end{equation}
so the instanton satisfies the equation of motion
\begin{equation}
\ddot{\phi} + \frac{3}{\rho} \dot{\phi} = \frac{d V}{d \phi}. \label{eq:eom}
\end{equation}
At the origin $\phi(0) \sim C$ and $\dot{\phi}(0) = 0$;  far from the bubble center $\lim_{\rho \rightarrow \infty} \phi(\rho) = A$.

\begin{figure}[h!] %  figure placement: here, top, bottom, or page
   \centering
   \includegraphics[width=3.6in]{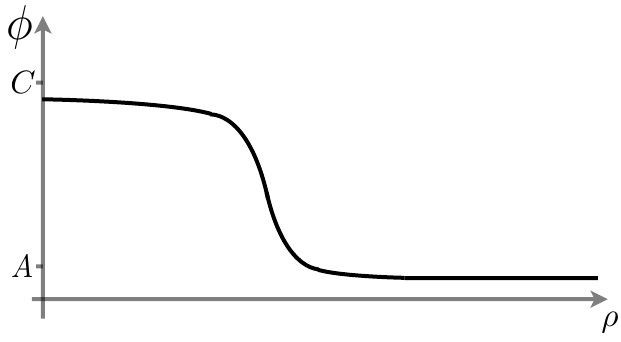}
   \caption{A cross-section through the center of the bubble. The field value interpolates from close to the true vacuum at $\rho = 0$ to exactly the false vacuum as $\rho \rightarrow \infty$. In the thin-wall limit this profile sharpens to a step function.} 
   \label{DIfig-profile}
\end{figure}

For the potential of Fig.~\ref{DIfig:potential} we know such a solution must exist by Coleman's undershoot/overshoot argument \cite{Coleman:1977py}. The argument starts by noticing that the equation for the bubble wall profile, Eq.~\ref{eq:eom},  is also the equation for a particle with position $\phi(\rho)$ sliding in the inverse potential, $-V(\phi)$, with friction that fades with time like $\rho^{-1}$.  For the desired solution, the particle starts at rest at time $\rho = 0$ near $C$, and asymptotically returns to rest as $\rho \rightarrow \infty$ balanced precariously atop the hill at $A$. We will use continuity to argue that such a solution exists. On the one hand, the particle undershoots $A$ if started too low on the hill. This is because it doesn't have enough energy to make it up the other side and instead settles down in the valley. On the other hand, the particle overshoots $A$ if started too high on the hill.  It takes such a long time to slide off the crest that friction is by then negligible; the particle has too much energy and blows on past to $\phi < A$. By continuity, there must be a critical value of $\phi(0)$ so that the particle comes to rest at $A$. %This value is associated with the instanton. 

\begin{figure}[h!] %  figure placement: here, top, bottom, or page
   \centering
   \includegraphics[height=2.5in]{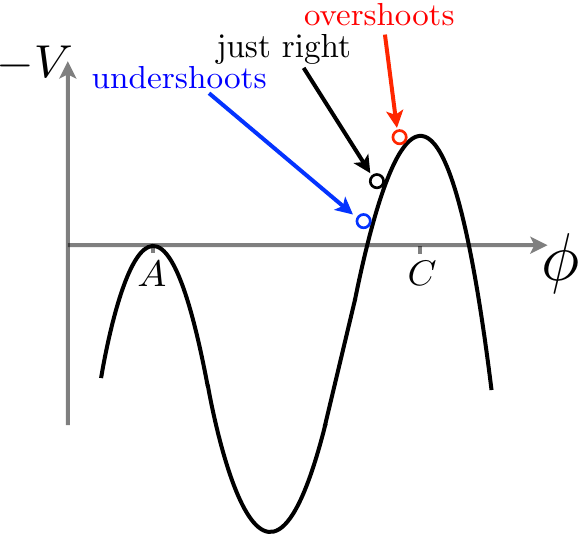}
   \caption{The undershoot/overshoot argument guarantees that an instanton solution  exists. A particle released too low undershoots and ends up at $\phi > A$. A particle released too high overshoots and ends up at $\phi < A$. By continuity, a particle released at the critical height asymptotically come to rest at $A$. This particle's trajectory gives the field profile of the instanton.}
   \label{fig:inversepotential}
\end{figure}

\pagebreak

It is sometimes useful to consider the limit in which the potential barrier separating the two vacua is tall and narrow. In this `thin-wall limit' the profile of Fig.~\ref{DIfig-profile} approximates a step function at a radius $\bar{\rho}$ given by 
 \begin{equation}
 \bar{\rho} = \frac{3 \sigma}{\epsilon},
 \end{equation}
where the domain wall tension is 
\begin{equation}
\label{eq:sigma}
\sigma \equiv \int_A^C d \phi \sqrt{2 V(\phi)}, 
\end{equation}
 and $\epsilon \equiv V_A - V_C$ is the difference in  energy density  between the two vacua.  This critical radius $\bar\rho$ is determined by a balance of forces.  The pressure differential gives the wall an outward push equal to $\epsilon$, and the surface tension gives the wall an inward pull equal to $3\sigma/\rho$.  The location of the bubble wall is thus at a local maximum of $S(\rho)$, an unstable fixed point.  Changing the radius of the bubble is the instanton's one and only negative mode.
 In the thin-wall limit, $\Delta S$ is given by 
 \begin{equation}
\Delta S  =  \frac{27 \pi^2 }{2} \frac{\sigma^4 }{\epsilon^3}. \label{eq:thinwallB}
\end{equation}

\section{Simple Examples of Abrupt Disappearance} \label{sec:simple}

In this section, we study in detail two examples of instanton disappearance.
 In Sec.~\ref{subsec:twostep} we look at abruptly disappearing instantons in a single-field potential, caused by an intervening minimum.   Such chains of minima are common in axion landscapes \cite{Abbott:1984qf}. 
In Sec.~\ref{subsec:runaway} we look at  abruptly disappearing instantons in a multi-field potential, caused by a runaway direction.  Runaway directions are common in landscapes that arise in extra-dimensional models \cite{Dine:1985he}. The non-existence of some instantons in runaway potentials was first noticed in \cite{Cvetic:1994ya}.

\subsection{Intervening Minimum} \label{subsec:twostep}

The potential of Fig.~\ref{fig:abruptpotential} has a false vacuum at $A$, an intermediate vacuum at $B$, and a true vacuum at $C$. We restrict ourselves to $V_A > V_B > V_C$, and, at first, to the thin-wall limit. Let's count the number of instantons this potential supports as we smoothly lower $V_B$ from $V_A$ towards $V_C$.  

\begin{figure}[h!] %  figure placement: here, top, bottom, or page
   \centering
   \includegraphics[width=\textwidth]{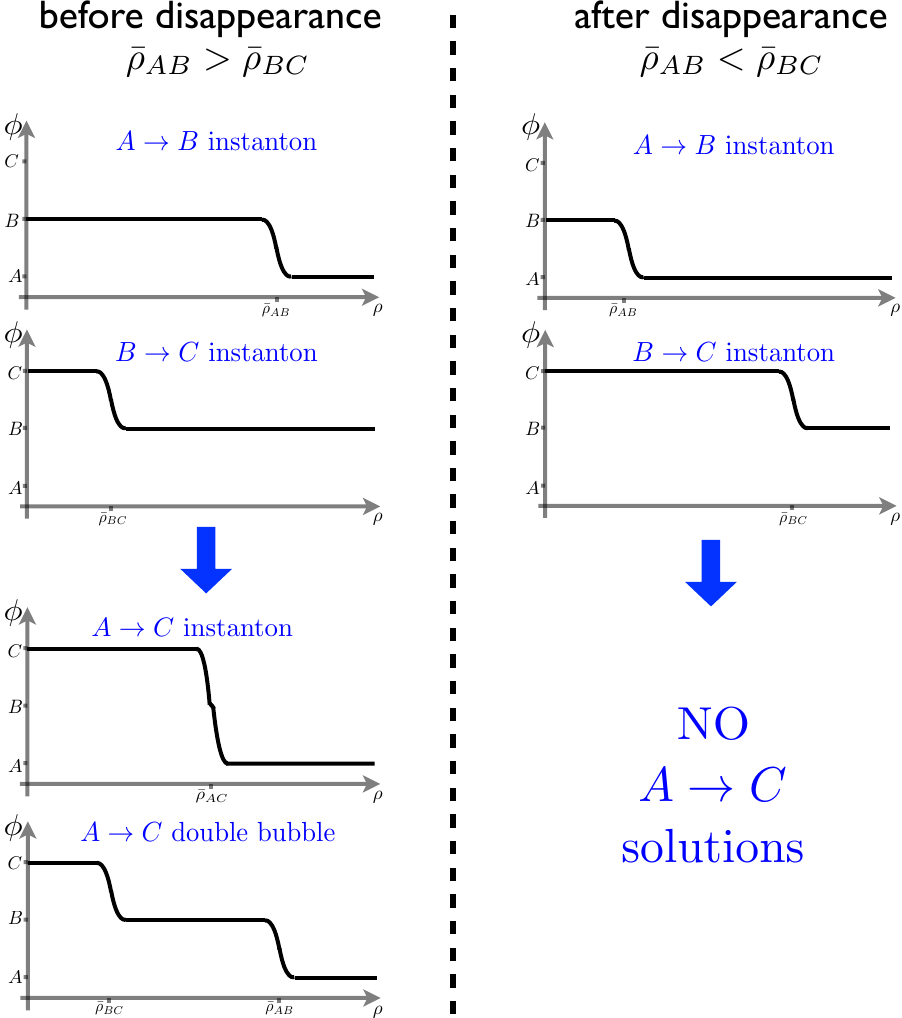} 
   \caption{There are two solutions that always exist: the $A B$ instanton (shown in both the top left and top right panes) and the $B  C$ instanton (shown in both the second left and second right panes). When $\bar{\rho}_{AB} > \bar{\rho}_{BC}$, there are two more solutions. The $AC$ instanton (shown in the third left pane) has a single wall that mediates directly from $A$ to $C$. The double bubble solution (shown in the bottom left pane)  is the concatenation of the $AB$ and $BC$ solutions.    This solution has an extra negative mode; if you flow down this mode in the direction of merging walls you reach the $AC$ instanton. When $\bar{\rho}_{AB} < \bar{\rho}_{BC}$, neither solution exists. }
      \label{fig:twostep}
\end{figure}

There are two instantons that  always exist. One is the $AB$ instanton, describing a bubble of $B$ embedded in $A$, which has a single negative mode associated with changing the radius of the bubble wall at $\bar{\rho}_{AB} \equiv 3 \sigma_{AB} / \epsilon_{AB}$. The other is the $BC$ instanton, which likewise has a single negative mode associated with changing $\bar{\rho}_{BC} \equiv 3 \sigma_{BC}/ \epsilon_{BC}$. As we lower $V_B$, $\bar\rho_{AB}$ shrinks in from infinity, $\bar\rho_{BC}$ grows outwards towards infinity. At a critical value of $V_B$, they pass. The profiles for these instantons are plotted in the top panes of Fig.~\ref{fig:twostep}. 
\pagebreak

\pagebreak
{\bf The $AC$ instanton.}  When $V_B$ is only a little less than $V_A$, there is also another instanton, which we'll call the $AC$ instanton.   This describes a bubble of $C$ directly embedded in $A$ with a single thin wall that spends negligible time in the $B$ vacuum.  This solution has one negative mode associated with changing its radius 
\begin{equation}
\bar{\rho}_{AC} =  \frac{3 \sigma_{AC}}{\epsilon_{AC}}=\frac{3 (\sigma_{AB} + \sigma_{BC})}{\epsilon_{AB} + \epsilon_{BC}}.
\end{equation}
This $AC$ instanton is plotted in the third left pane of Fig.~\ref{fig:twostep} and has action 
\begin{equation}
(\Delta S)_{AC} = \frac{27 \pi^2}{2} \frac{\sigma_{AC}^{\,4}}{\epsilon_{AB}^{\,3}}=\frac{27 \pi^2}{2} \frac{( \sigma_{AB} + \sigma_{BC} )^4}{(\epsilon_{AB} + \epsilon_{BC})^3}. 
\end{equation}
For $V_B$ only a little less than $V_A$, the $A \rightarrow B$ decay is slow, so the $AC$ instanton is the fastest decay mode of $A$. As $V_B$ is lowered, the $A \rightarrow B$ decay gets faster and eventually takes over as the fastest decay. Lower $V_B$ a little further and something more dramatic happens: the $AC$ instanton abruptly disappears. 

\begin{figure}[h]
   \centering
   \includegraphics[width=\textwidth]{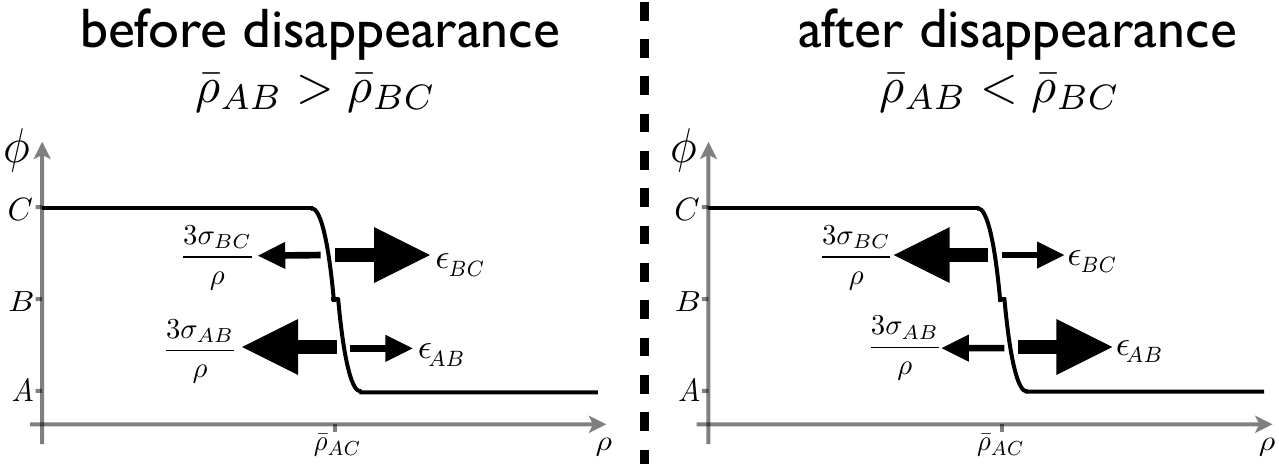} 
  \caption{Before disappearance, the $BC$ segment of the wall feels a net outwards push, the $AB$ segment of the wall feels a net inwards pull, and they are jammed together. After disappearance,  the net forces reverse and the wall disintegrates. } 
  \label{fig:wallintegrity}
\end{figure}

The $AC$ instanton exists whenever 
\begin{equation}
\bar{\rho}_{AB} \geq \bar{\rho}_{BC}, \label{eq:radiusinequality}
\end{equation}
but lower $V_B$ further and the $AC$ instanton's wall falls apart. As shown in Fig.~\ref{fig:wallintegrity}, the $AB$ segment of the wall is subject to an outwards force per unit area of $\epsilon_{AB}$ caused by the pressure differential, and an inwards force per unit area of $3 \sigma_{AB} / \rho$ caused by the surface tension. So long as the net force on the $AB$ segment is inwards, and the net force on the $BC$ segment is outwards, they're jammed together and the wall coheres. But  try to construct a wall when Eq.~\ref{eq:radiusinequality} is not satisfied and the wall is torn apart.
 
At the moment of disappearance, when Eq.~\ref{eq:radiusinequality} is saturated, we have $(\Delta S)_{AB} < (\Delta S)_{AC} < \infty$, so we have the \emph{abrupt} disappearance of a \emph{subdominant} instanton. As promised in the introduction, the instanton is the subdominant decay mode out of $A$ by the time it abruptly disappears. But what annihilates it?

{\bf The double bubble solution.} Whenever the $AC$ instanton exists, there is also another solution, the `double bubble' solution, which is the concatenation of  the $AB$ and $BC$ instantons: it's a bubble of $C$ sitting inside a bubble of $B$ in a background of $A$. This solution has action $S - S_A = (\Delta S)_{AB}  + (\Delta S)_{BC} \geq (\Delta S)_{AC}$, and has \emph{two} negative modes: it inherits one from each of its constituents.  Perturbing the position of either wall  reduces the action.  In particular, perturbing the two walls towards each other monotonically lowers the action until they meet at $\bar{\rho}_{AC}$, so that the extra negative mode of the double bubble solution corresponds to deforming it into the $AC$ instanton. 

We have seen that lowering $V_B$ past the point where $\bar\rho_{AB} = \bar\rho_{BC}$ makes the $AC$ instanton cease to exist. It also makes the double bubble cease to exist because the two bubbles just don't fit together anymore. 
As we smoothly alter the parameters of the potential, the two solutions---an instanton, and a Euclidean solution with higher action and an extra negative mode---merge, annihilate and disappear.  

\vspace{-.31cm}
\begin{equation}\nonumber
\star \, \star\,\star
\end{equation}

\begin{figure}[htbp] %  figure placement: here, top, bottom, or page
   \centering
   \includegraphics[width=\textwidth]{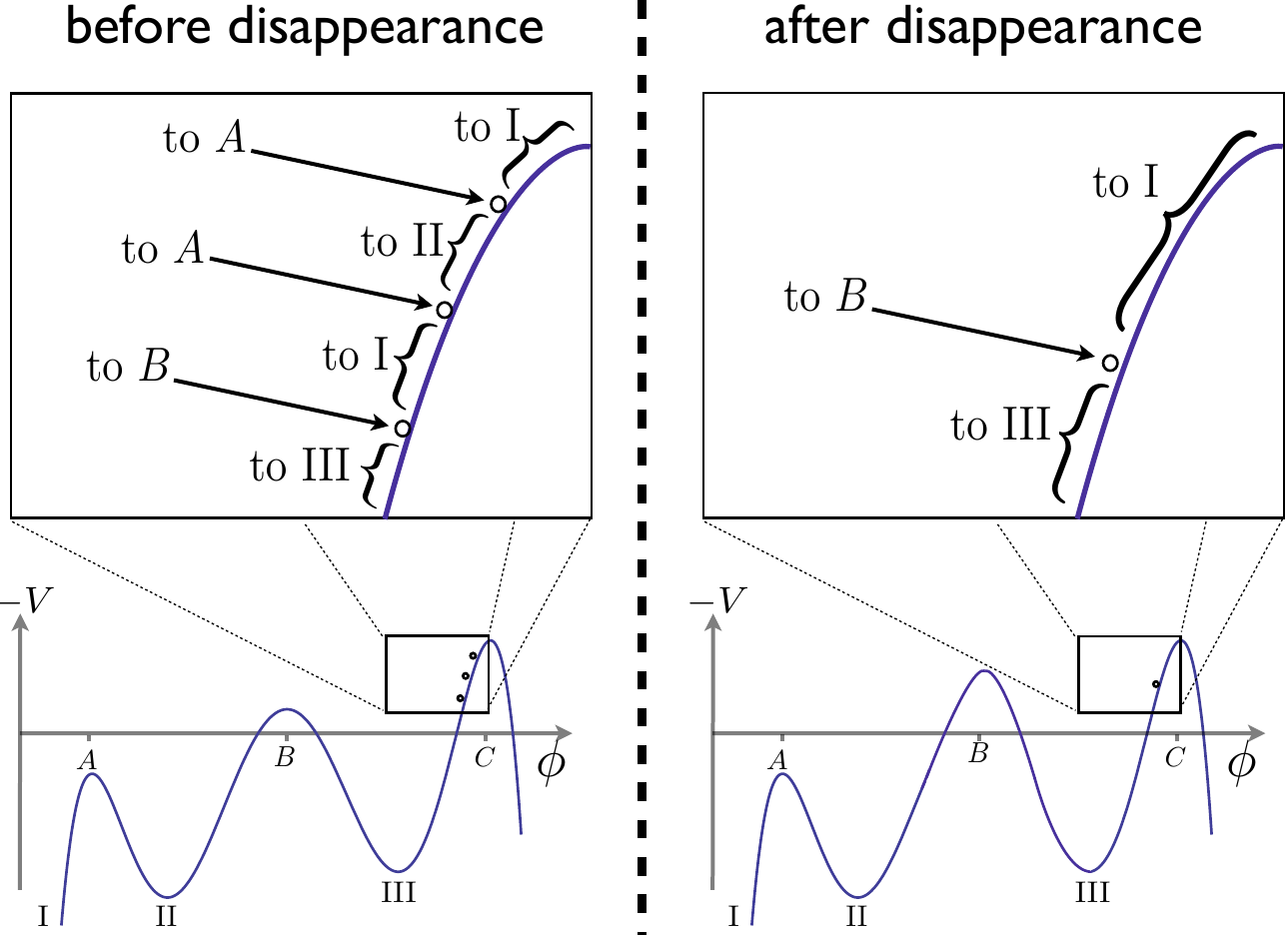} 
   \caption{Where the particle ends up as a function of where it is released on the inverse potential.   Critical points that end on $A$ separate `to I' regions from `to II' regions, so there must be an even number of them. }
   \label{fig:upsidedown2}
\end{figure}

Away from the thin-wall limit, 
the story is the same.
By extending the undershoot/over-shoot argument to this potential, we will see that it no longer guarantees the existence of an $A \rightarrow C$ instanton, but it \emph{does} guarantee that when there is one $A \rightarrow C$ solution, there must be a second. Figure~\ref{fig:upsidedown2} illustrates the two cases, before and after disappearance.  We begin by nothing the features the two cases have in common. 

Let's look low on the hill. Start the particle too low and it languishes in III. Above that, there is a critical starting point so that the particle comes to rest on $B$. Infinitesimally higher still and the particle spends a long time crossing the $B$ peak before sliding off; during that long time, the friction term  becomes negligible, so the particle blows past $A$ and on to I. 

Now let's look high on the hill. As before, there is a region of release points near the crest that causes the particle to overshoot everything and end in I. 
In summary, in both cases there's a region high on the hill that causes the particle to end in I, and a region low on the hill that causes the particle to end in I. But there may be intermediate release points that lead to $A$. 

Suppose there is a starting point that makes the particle comes to rest on $A$. Perturb the starting point one way, and the particle  approaches $A$ with just too much velocity; it continues past and is carried away towards I.  Perturb the starting point the other way, and the particle approaches $A$ with not quite enough velocity; it turns around and eventually comes to rest in II.  A point that leads to $A$ must therefore lie between a region that leads to I and a region that leads to II.  This means that between the region low on the hill that leads to I and the region high on the hill that leads to I, there must be an \emph{even} number of critical points that lead to $A$; there can be two, as there are before disappearance, or zero as there are after disappearance. 

\vspace{3cm}

\pagebreak

\subsection{Runaway} \label{subsec:runaway}
Consider two (3+1)-dimensional fields, $x(\vec{r},t)$ and $y(\vec{r},t)$, that live in a potential
\begin{equation}
V(x,y)=-y
\end{equation}
except for two narrow, essentially point-like minima
\begin{eqnarray}
V(x=0, y=0)  & = & E \ \ < 0 ,   \ \ \  \, \, \, \hspace{1pt} \ \textrm{ the false vacuum, and} \nonumber \\
V(x = \delta, y=0) & = & E - \epsilon \ \  \ \, \ \ \,  \ \ \ \   \ \textrm{the true vacuum}.
\end{eqnarray}
The potential  resembles a tilted gold course, with two holes along the $x$-axis. A field homogeneously in the false vacuum has two decay directions: it can decay to the true vacuum at $(\delta,0)$, or it can can decay out in the $y$-direction. In this subsection, we will consider the first decay and show that as $\delta$ is smoothly increased the corresponding instanton abruptly disappears.
\begin{figure}[htbp] %  figure placement: here, top, bottom, or page
   \centering
   \includegraphics[width=2.7in]{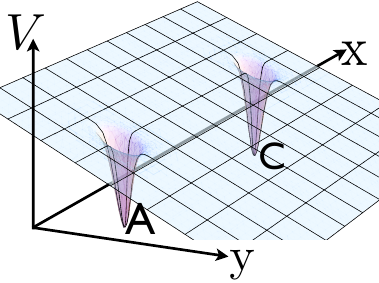} 
   \caption{The potential, $V(x,y)$, as a function of the two fields, $x$ and $y$, resembles a tilted golf course.  If the minima are moved too far apart, the instanton that connects them disappears.}
   \label{fig:tilted}
\end{figure}

In the thin-wall limit, the field profile of the bubble wall is given by the path through field space that minimizes the wall tension $\sigma$, defined by Eq.~\ref{eq:sigma}. This optimal path is determined by a tradeoff: straight paths that follow $y=0$ are shorter, while loopy paths that detour to large $y$ are longer, but experience a lower potential. 

We have seen that there is a mathematical equivalence between minimizing the wall tension and solving for the trajectory of a frictionless particle sliding in the inverse potential, $-V(x,y)$. In the analog problem, the particle leaves the false vacuum spike with an initial velocity that is fixed by the false vacuum energy, $v=\sqrt{-2E}$, and must be aimed so as to land precisely on the true vacuum. The only adjustable parameter is the angle that the trajectory makes with the $x$-axis as it leaves the spike; this angle $\theta$ must be chosen so that the particle arrives on target.  The 
 particle covers
\begin{equation}
\Delta x=2 v^2 \sin \theta \cos \theta = -2E \sin 2 \theta 
\end{equation}
before reaching $y=0$ again; 
in order to land on the true vacuum, we require that $\Delta x=\delta$. 
For short distances ($\delta<-2E$), there are two ways to do this: a `low road' with $\theta < 45 ^\circ$, and a `high road' with $\theta > 45 ^\circ$. 
For long distances ($\delta > - 2 E$), there is no way to do this: the maximum distance a frictionless particle can cover is achieved at $\theta = 45^{\circ}$. 

 \begin{figure}[htbp] %  figure placement: here, top, bottom, or page
    \centering
    \includegraphics[width=2.8in]{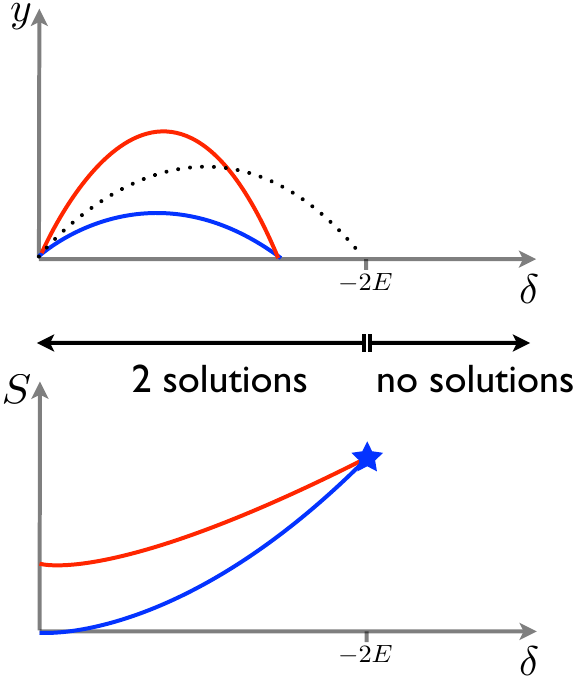} 
    \caption{There are two ways to reach any $\delta < {-2E}$: the small-$\theta$ low road, in blue, and the large-$\theta$ high road, in red. The low road has  lower action and only one negative eigenmode; the high road has higher action and an extra negative mode. As $\delta \rightarrow {-2E}$, $\theta \rightarrow 45 ^{\circ}$ from either side, and the two solutions merge and annihilate. For $\delta > {-2E}$ there are no solutions. }
    \label{Sofdelta}
 \end{figure}

\begin{figure}[htbp] %  figure placement: here, top, bottom, or page
    \centering
    \includegraphics[width=\textwidth]{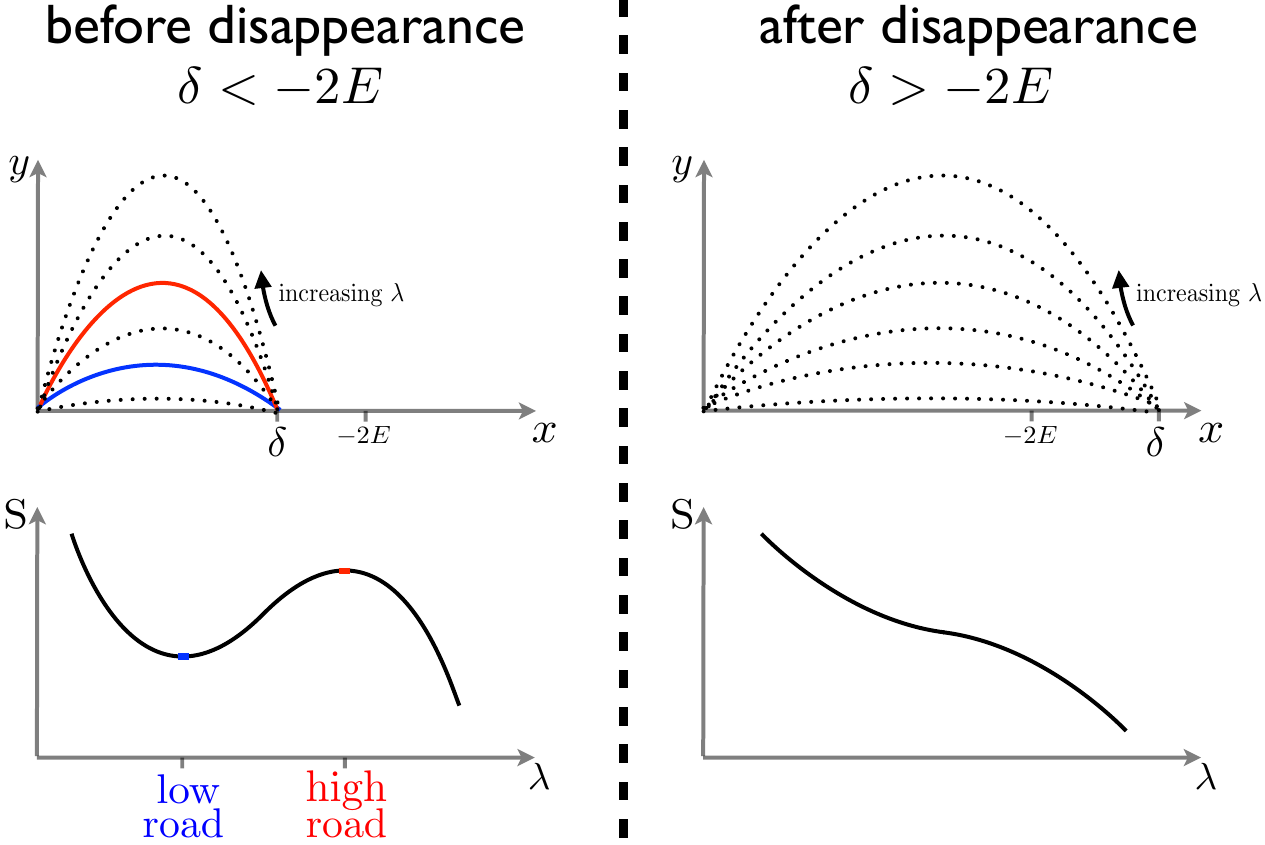} 
    \caption{The action for the one-parameter family of solutions described by $y = \lambda x (\delta - x)$, for a given $\delta$. For $\delta < -2E$, the action has a minimum (the low road), and a maximum (the high road). As $\delta \rightarrow -2E$, the maximum and minimum merge to form an inflection point. For $\delta > -2E$, there are no stationary points.}
    \label{sigmaofa}
 \end{figure}

The parabolic trajectories can be plugged into the WKB formula, Eq.~\ref{eq:sigma}, to give
\begin{eqnarray}
\sigma_{\textrm{high road}} & = & \frac{\sqrt{2}(-2E)^\frac{3}{2}}{3} \left[ 1 + 3 \frac{\delta^2}{(-2E)^2} + \left( 1 - \frac{\delta^2}{(-2E)^2} \right)^{\frac{3}{2}} \right]^{\frac{1}{2}}, \textrm{ and}  \\
\sigma_{\textrm{low road}} & = & \frac{\sqrt{2} (-2E)^\frac{3}{2}}{3} \left[ 1 + 3 \frac{\delta^2}{(-2E)^2} - \left( 1 - \frac{\delta^2}{(-2E)^2} \right)^{\frac{3}{2}} \right]^{\frac{1}{2}}.
\end{eqnarray}
The low road has lower tension, and hence lower action. The high road, by contrast, is the road less travelled by. Both solutions have the standard negative mode associated with \emph{moving} the bubble wall, but the high road has an extra negative mode associated with \emph{deforming} the bubble wall. Consider the family of wall profiles given by $y=\lambda x(\delta-x)$. As shown in Fig.~\ref{sigmaofa}, the low road is a local minimum of the action under this deformation, but the high road is a local maximum---it can lower its action by flowing either down to the low road, or out to $y = \infty$.

As we smoothly deform $\delta$ towards $-2E$, the two $\theta$s approach $45^{\circ}$ from either side. At the critical point  the low and high roads approach, merge and annihilate---a minimum of $S(\lambda)$ meets a maximum of $S(\lambda)$ and becomes an inflection point. For larger $\delta$, $S(\lambda)$ has no extrema: \emph{any} wall that interpolates between the two vacua will slide off the potential and slide away to $y\rightarrow\infty$.  It's not just that there's no tunneling instanton between the vacua, it's more than that: even if you make `by hand' a bubble of $C$, then the region of $C$ won't expand---instead the wall thickens and slides down the slope, expanding in both directions and consuming true and false vacuum alike \cite{Aguirre:2009tp}.  What seemed like a transition to $C$ is actually a transition in the runaway direction.

By the time the low road and high road meet, both have higher action than the instanton that mediates tunneling out towards $y=\infty$: we again have the \emph{abrupt} disappearance of a \emph{subdominant} instanton. 

\vspace{1.3cm}

\pagebreak
\section{Discussion}
\label{sec:summary}

In the last section, we studied  two examples of disappearing instantons, and showed that they exhibit three common characteristics.  In this section, we see how disappearing instantons arise in the semiclassical approximation to the path integral. 

\begin{figure}[htbp] %  figure placement: here, top, bottom, or page
   \centering
   \includegraphics[width=4.1in]{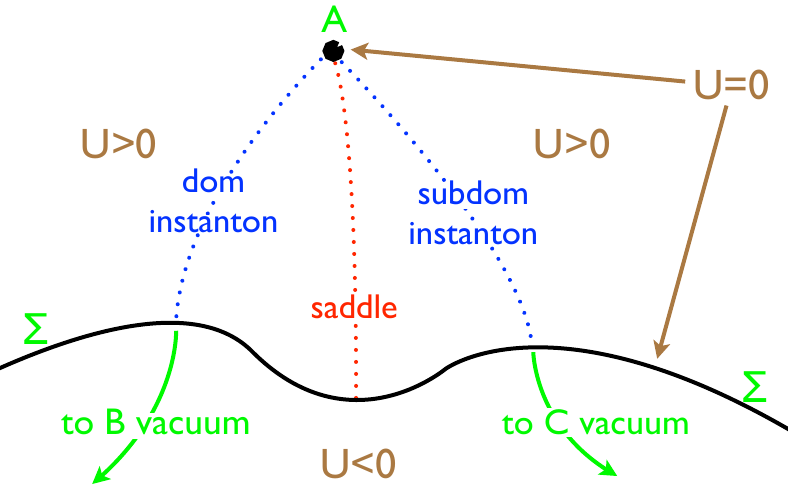} 
   \caption{The false vacuum $A$ has $U=0$ and is surrounded by configurations with $U>0$.  Since the false vacuum is unstable, there are distant configurations with $U<0$.  In between the region with $U>0$ and the region with $U<0$, there is a boundary, $\Sigma$, the set of configurations that, like the false vacuum, have $U=0$.  Of the paths that leave $A$ and end on their first intersection with $\Sigma$, the global minimum of the action is the dominant instanton.  Local minima of the action are subdominant instantons and saddles are solutions with extra negative modes.  Different regions on $\Sigma$ may classically evolve to different true vacua.}
   \label{fig:pathtosigma}
\end{figure}

Instantons are paths through configuration space that extremize the Euclidean action $S$.  
They mediate tunneling because they traverse a classically forbidden region of the potential.  They start on the false vacuum and end on a state with equal energy that will classically evolve to the true vacuum; in between, the potential  energy is always bigger than at either of the two endpoints. The  space of possible  instanton endpoints is a codimension-one surface of configurations that  have the same potential energy as the false vacuum (let us say zero energy); we will call this surface $\Sigma$. 
  Figure~\ref{fig:pathtosigma} shows three paths that leaves the false vacuum and ends on $\Sigma$.

For quantum particle mechanics, configuration space is the space of positions, $x$. For quantum field theory, configuration space is the space of field configurations defined on a spatial slice, $\phi(\vec{x})$.  The potential $U$ is defined over configuration space and, for a standard scalar field theory, it is
\begin{equation}
U[\phi(\vec{x})] = \int d^3 x \left[ \frac{1}{2} \left(\vec{\nabla} \phi(\vec x)\right)^2  + V\left(\phi(\vec x)\right) -V_A\right] .
\end{equation}
For example, the familiar Euclidean bubble instanton, sliced on constant $\tau$ surfaces, is such a path through configuration space.  It starts on the false vacuum $\phi(\vec{x},\tau \rightarrow -\infty)$ with $U=0$ and  it ends on a critical bubble $\phi(\vec{x},\tau=0)$. The critical bubble also has $U=0$ and so is an element
 of $\Sigma$.
 Intermediate $\tau$ slices give intermediate configurations along the path, and all have $U>0$.\footnote{In quantum particle mechanics, $\Sigma$ does not have to be connected, but in quantum field theory in Minkowski space, it does.  For instance, we can smoothly deform a critical bubble of $B$ into a critical bubble of $C$ while keeping $U=0$. First make the $B$ bubble bigger while making a new bubble of $C$ a great distance away, then make the $C$ bubble bigger while removing the bubble of $B$.  Subcritical bubbles make a positive contribution to $U$, so to compensate either for making the $C$ bubble or for removing the $B$ bubble, we must simultaneously make the other bubble supercritical, so that it makes a negative contribution to $U$.  Because there is infinite volume, the two bubbles can be made far enough apart that they do not interact and made big enough to compensate for any potential barrier.  However, even though $\Sigma$ is connected, different states in $\Sigma$ can classically evolve to different true vacua.}

Constrained to paths that start in the false vacuum and end on their first intersection with $\Sigma$, the Euclidean action is necessarily positive: $S$ is a sum of positive kinetic energy and positive $U$.  The action is bounded from below, and the global minimum of the action is the dominant instanton mediating decay out of the false vacuum.  All perturbations to this path that keep the endpoint on $\Sigma$ necessarily increase $S$.  Coleman explained in \cite{Coleman:1987rm} that the instanton's single negative mode comes from moving the endpoint of the path off $\Sigma$. This either shortens the path, or lengthens it, but lengthens it into the region of negative $U$, in either case decreasing $S$. For example, in the thin-wall limit, changing the radius of the bubble moves the endpoint off $\Sigma$.

Besides this global minimum, $S$ can have other extrema: local minima are instantons, and saddles are solutions with extra negative modes.  We're interested in what happens to minima under smooth changes in the potential. We have identified two possibilities, shown in Fig.~13.

\begin{figure}[h!] %  figure placement: here, top, bottom, or page
   \centering 
   \includegraphics[width=6in]{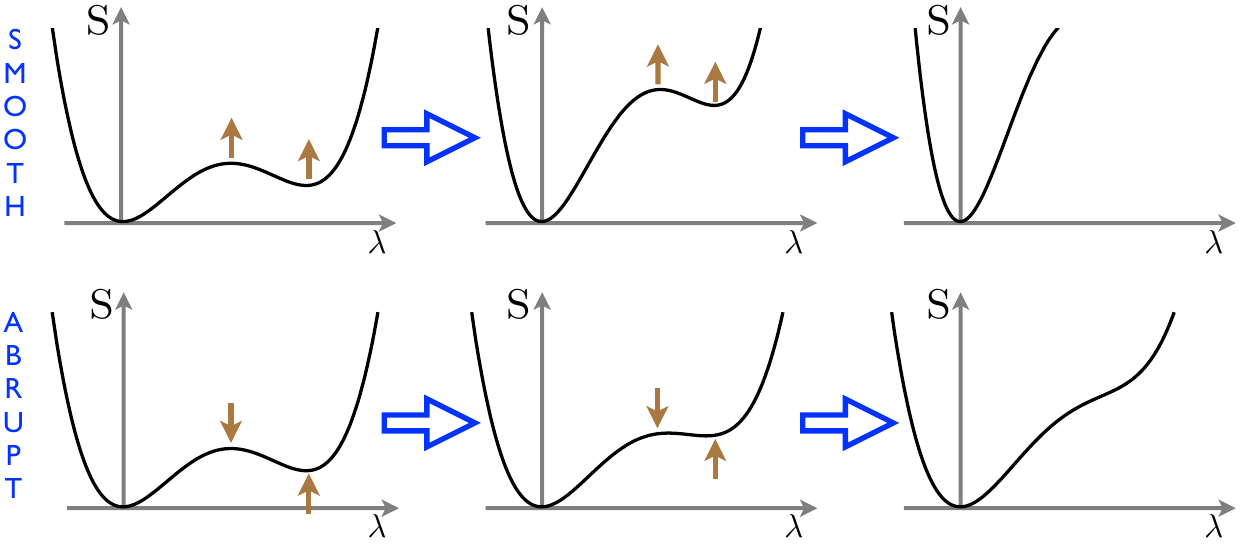}  \label{fig:abruptsmoothdisappearances}
   \caption{The Euclidean action $S$ is plotted as a function of $\lambda$, which parameterizes a family of paths that contains two instantons.  As the potential is changed, we have seen two ways an instanton can disappear.  In a smooth disappearance, a local minimum lifts up to $S\rightarrow\infty$, carrying a neighboring saddle point with it. In an abrupt disappearance, a local minimum and a saddle annihilate.}
\end{figure}

First, a local minimum can lift off to $S\rightarrow\infty$, which corresponds to a smooth disappearance.  If a subdominant instanton smoothly disappears, then it must take with it neighboring saddle points.  (For instance, in the potential of Fig.~\ref{fig:abruptpotential}, as the $B$ vacuum is raised towards $A$, both the $AB$ instanton and the double bubble solution smoothly disappear together.)  The global minimum, which is the dominant instanton, can also smoothly disappear leaving the  vacuum completely stable.

Second, a local minimum can annihilate with a saddle to become an inflection point \cite{Maxwell}, which corresponds to an abrupt disappearance.   The annihilating solution has a single extra negative mode, for a total of two negative modes, the one it shares with the instanton associated with moving the endpoint off $\Sigma$ and the other associated with deforming the solution into the instanton. Because an inflection point leads in one direction to lower $S$, only subdominant instantons can abruptly disappear.\footnote{One might worry that there is another possible fate for instantons. The space of tunneling paths has a boundary because we have restricted attention to paths that end on their \emph{first} intersection with $\Sigma$. The concern is that, by smoothly changing the parameters, the instanton can be pushed over the boundary, can be made to intersect with $\Sigma$ before it arrives at its original escape point. But this cannot happen: in fact the boundary repels instantons. We will argue by contradiction. As we smoothly change the potential, consider the instanton just before it is pushed over the boundary: it almost prematurely intersects with $\Sigma$, but not quite. Instantons are solutions to the equations of motion with the same energy as $\Sigma$, which means they  must approach and leave $\Sigma$ perpendicularly. So the not-quite-prematurely-intersecting instanton must, after its near-miss, continue back along the path it came, returning to $A$, rather than going on to some other part of $\Sigma$. But this contradicts our assumption that it was a tunneling instanton.}

This discussion has taken place in the context of the zeroth order semiclassical approximation. We have shown that the exponent in $\Gamma = \textrm{prefactor} \times  e^{- \Delta S / \hbar} $ stays finite up to the moment of abrupt disappearance. 
This exponent is determined by the instanton, and the prefactor gives the contribution of neighboring paths. The dominant contribution to the prefactor comes from paths with action within a few $\hbar$ of the instanton, but these paths only count if they still mediate decay to $C$. As the moment of annihilation is approached, these neighboring paths start spilling over the maximum; they don't evolve to $C$, and instead slip off in the dominant decay direction.  The prefactor therefore goes to zero, softening the disappearance, with resolution $O(\hbar)$.  The disappearance is extremely, but not infinitely, sharp.
(The standard one-loop approximation to the prefactor misses this effect.  It treats the action as quadratic in perturbations around the instanton, ignoring higher-order terms, so that the prefactor is given by Gaussian integrals and goes as $[\det(S'')]^{-1/2}$, as was shown in \cite{Callan:1977pt}.  At disappearance, the action has vanishing second derivative, so it is unreasonable to ignore higher-order terms and the approximation breaks down: in fact the `approximation' diverges, whereas the prefactor actually goes to zero.)

Events in which stationary points of a function abruptly disappear are classified by catastrophe theory \cite{Arnold}.  The simplest case, where a minimum and a maximum annihilate, is called a fold catastrophe; in cases with enhanced symmetry, stationary points can annihilate in higher-order catastrophes: we'll see an example in the next section.

\section{Disappearance of Gravitational Instantons} \label{sec:complex}

In this section, we  examine the disappearance of gravitational instantons.  We see that the characteristics we observed in a non-gravitational context carry over to the examples we consider here, despite the complications of gravity.
 We see that the `gravitational blocking' effect of \cite{Coleman:1980aw} is a smooth disappearance; that the abrupt disappearance of an instanton due to an intervening minimum exhibits the same three general characteristics  it did in the non-gravitational case; and that the enhanced symmetry of the Hawking-Moss instanton gives rise to a higher-order catastrophe.  

\subsection{Review of Thin-wall Gravitational Instantons}
Gravitational instantons were first described by Coleman and De Luccia \cite{Coleman:1980aw}.  Spacetime may now be curved, but the instantons are still assumed to be O(4)-symmetric so the metric is
\begin{equation}
ds^2=d\rho^2+a(\rho)^2d\Omega_3^{\,2}.
\end{equation}  
Neglecting gravity, a thin-wall bubble is completely specified by $\sigma$ and $\epsilon$. With gravity, the absolute value of the potential is also important.  

\begin{figure}[h!] %  figure placement: here, top, bottom, or page
   \centering
   \includegraphics[width=5in]{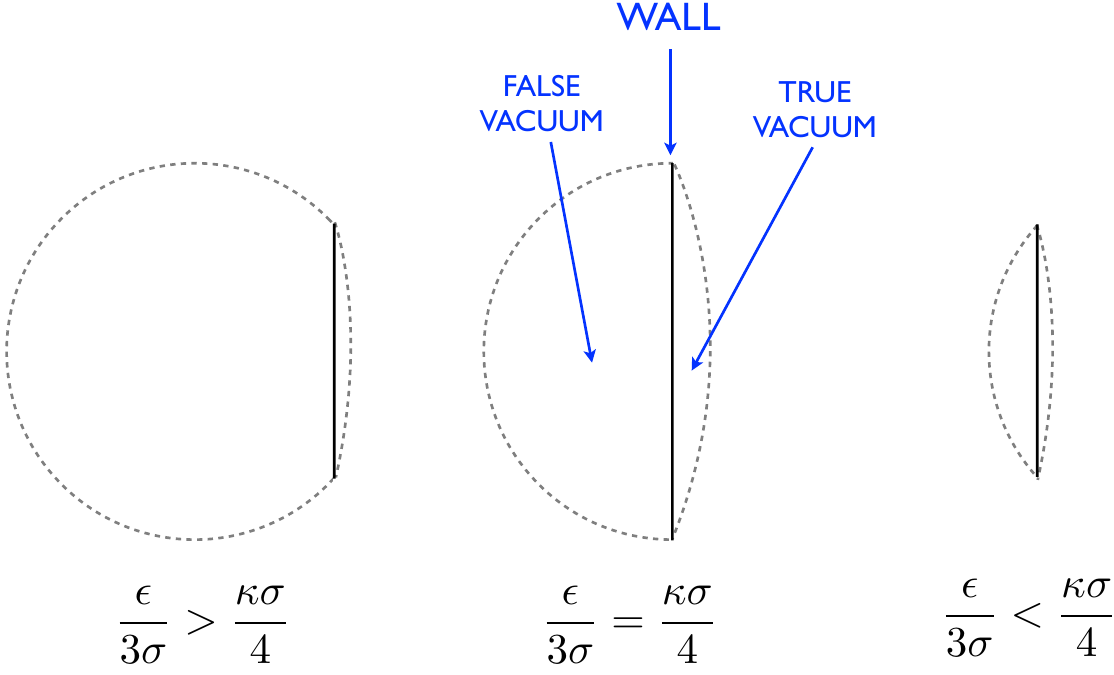} 
   \caption{For decays from de Sitter, a thin-walled instanton has a less than hemispherical section of true vacuum conjoined with a section of false vacuum. As $\sigma$ is increased, ever less of the false vacuum is left over.}
   \label{fig:typeAtypeB}
\end{figure}

Euclidean de Sitter space is a four-sphere with curvature radius $\ell = \sqrt{3 / \kappa V}$.  The thin-walled instanton will glue together a (less than hemispherical) section of true vacuum with a section of false vacuum. If $\epsilon/ 3 \sigma  > \kappa \sigma/4$, there is more than a hemisphere of false vacuum; otherwise there is less than a hemisphere, as depicted in Fig.~\ref{fig:typeAtypeB}. 
(If a vacuum is AdS, with $V<0$, then the corresponding $\ell$ is imaginary, and rather than a section of a sphere, there is a section of a hyperboloid.) The thin wall lies at $a(\bar \rho)$, which can be obtained by using the junction condition \cite{Coleman:1980aw}
\begin{equation}
\label{eq:rhobargrav}
\frac{1}{a(\bar{\rho})^2} = \frac{1}{\ell_{\text{f}}^2} + \left(  \frac{\epsilon}{3 \sigma} - \frac{\kappa \sigma }{4} \right)^2 = \frac{1}{\ell_{\text{t}}^2}  + \left(  \frac{\epsilon}{3 \sigma} + \frac{\kappa \sigma }{4}  \right)^2.
\end{equation}

\subsection{Smooth disappearance: gravitational blocking} \label{sec:gravitationalblocking}
We have seen how instantons mediating decays in a fixed Minkowski space smoothly disappear as the true vacuum is raised to be degenerate with the false.  As $\epsilon \rightarrow 0$, both the bubble radius and the Euclidean action diverge.
The gravitational instanton that mediates decay from Minkowski to AdS also smoothly disappears as the AdS true vacuum is raised, only now it disappears \emph{before} they become degenerate. Equation~\ref{eq:rhobargrav} shows that $a(\bar{\rho}) \rightarrow \infty$ (and consequently $\Delta S\rightarrow \infty$) as 
\begin{equation}
\ \ \ \ \ \ \ \ \ \ \ \ \ \ \ \ \ \ \ \ \frac{\epsilon}{3 \sigma} - \frac{\kappa \sigma}{4} \rightarrow 0 \ \ \ \ \ (\textrm{or } \sqrt{-3/\kappa V_f} \textrm{ for an AdS false vacuum}) .
\end{equation}

An interpretation of this behavior was given by Coleman and De Luccia \cite{Coleman:1980aw}.  Quantum decay from Minkowski produces a zero-energy critical bubble, where the negative energy of the interior true vacuum compensates for the positive energy of the surface.  But when the interior is negatively curved, the surface area to volume ratio is bounded from below and even an infinitely large bubble may still have positive energy. The instantons are gravitationally blocked \cite{Coleman:1980aw, Aguirre:2006ap}.

Gravitational blocking happens only for Minkowski or AdS false vacua. Since Eq.~\ref{eq:rhobargrav} implies that instantons from de Sitter are always compact, and since Euclidean de Sitter false vacua are themselves always compact, $\Delta S$ is necessarily finite. From de Sitter space there are no smooth disappearances.

\subsection{Abrupt disappearance: intervening minimum}

In Sec.~\ref{subsec:twostep} we described the disappearance of an instanton joining two vacua, $A$ and $C$, due to the intervention of a third vacuum, $B$. We saw that as the instanton disappears, its action stays finite, and it is annihilated by another Euclidean solution with an extra negative mode. Here, we will see that the same story applies in the presence of gravity. 

As before, in order for there to be a double bubble solution, we must be able to concatenate the two bubbles. As in flat space, this means that $a(\bar{\rho}_{AB}) > a(\bar{\rho}_{BC})$; but we now have a second condition, that the $BC$ transition involve more than a hemisphere of $B$.  Taken together, these two conditions are equivalent to
\begin{equation}
 \frac{\epsilon_{BC}}{3 \sigma_{BC}} - \frac{\kappa \sigma_{BC} }{4} \geq \frac{\epsilon_{AB}}{3 \sigma_{AB}}  +  \frac{\kappa \sigma_{AB} }{4},
\end{equation}
which is the gravitational version of the force balance relation on the $AC$ wall. For high $V_B$ there is both a double bubble solution and an $AC$ instanton. As $V_B$ is lowered, the two solutions annihilate. (Notice that this condition depends only on $\epsilon$'s, not on $V$'s, and so is unchanged by shifting the whole potential.)

As in the non-gravitational case,  the $AC$ instanton has lower action than the double bubble solution 
\begin{equation}
\Delta S_{AC} \leq \Delta S_{AB} + \Delta S_{BC}.
\end{equation}
At the moment of disappearance,  this inequality is saturated so the disappearing instanton is subdominant to tunneling directly to $B$.

\vspace{-.31cm}
\begin{equation}\nonumber
\star \, \star\,\star
\end{equation}

Away from the thin-wall limit, the story is the same.  The field equation, the generalization of Eq.~\ref{eq:eom}, becomes 
\begin{equation}
\ddot{\phi}(\rho) + 3 H (\rho) \,  \dot{\phi}(\rho) = \frac{ d V } {d \phi} ,
\end{equation}
and the null energy condition requires $\dot{H} \equiv \partial_{\rho} (\dot{a}/a)  < 0$. Thus, like in the non-gravitational case, the `friction' is always becoming less positive (unlike in the non-gravitational case, eventually even becoming negative).  As a consequence, the same extension of the undershoot/overshoot argument again shows that solutions must appear and disappear in pairs. 

\subsection{Hawking-Moss and the Cusp}

In this subsection we see a new type of abrupt disappearance, associated with enhanced symmetry. The Hawking-Moss (HM) instanton represents a thermal transition in de Sitter space, in which the whole horizon volume homogeneously fluctuates to the top of the barrier. The instanton has maximal O(5) symmetry; everywhere on the four-sphere that is Euclidean de Sitter the field is uniformly perched on the crest  \cite{Hawking:1981fz}. By contrast, every other instanton breaks some symmetry: the Coleman-De Luccia (CDL)  instanton, for example, breaks O(5) down to O(4). The broken symmetry generators give the CDL solution its four zero modes; there's a continuum of rotated versions of the same solution.

\begin{figure}[htbp] %  figure placement: here, top, bottom, or page
  \centering
  \includegraphics[width=5.95in]{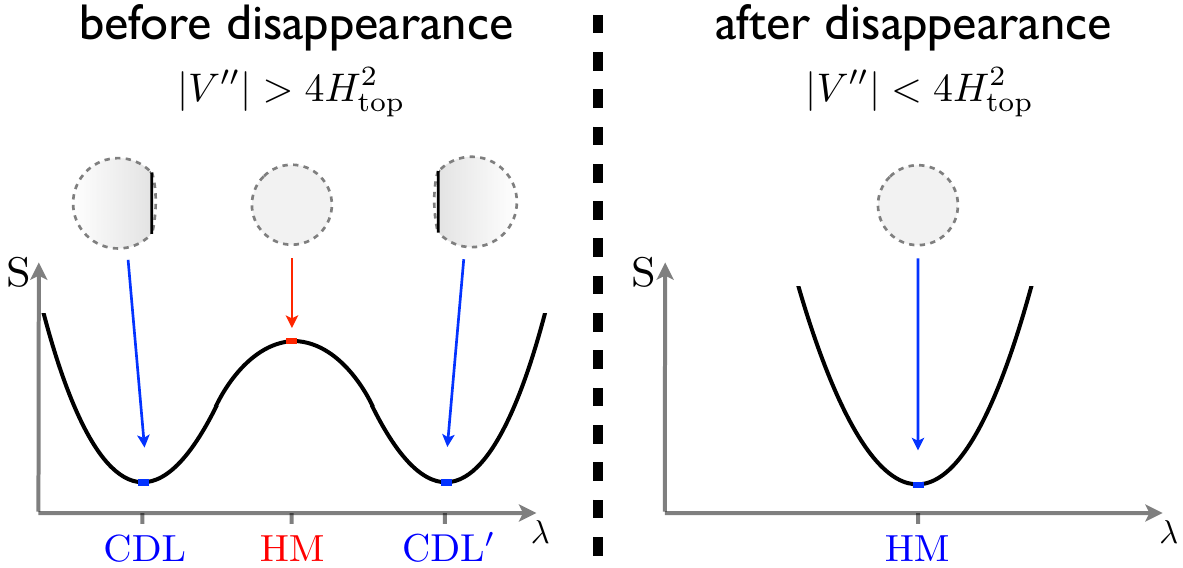}
  \caption{The Coleman-De Luccia instantons merging with the Hawking-Moss solution in a multi-dimensional cusp catastrophe. Before disappearance, the HM has five negative modes, and there is a basin of CDL instantons,  connected to each other by the four zero modes associated with rotating the solutions. These CDL instantons have a single negative mode. After disappearance, the HM instanton remains, but has only a single negative mode.}
  \label{fig:CDLHM}
\end{figure}

It was noticed in \cite{Jensen:1988zx}, that flattening the potential barrier makes the CDL solution disappear. 
For most potentials, as the potential gets flatter on top the bubble  wall gets thicker until, when $|V''| = 4 H_{\textrm{top}}^2$, the wall fills the entire horizon volume; the CDL instanton is all wall and has merged with the HM instanton.  As it merges, so too do the rotated CDL solutions, as in Fig.~\ref{fig:CDLHM}.  A ring of minima and a maximum annihilate into a single minimum: the HM swallows the CDL solutions, and in so doing changes its number of negative modes. This is a multidimensional version of the cusp catastrophe.

Let's count the negative modes of the HM. 
The $\ell=0$ spherical harmonic is always a negative mode---the field can lower its action by uniformly slipping off the top. The $\ell=1$ spherical harmonic modes are negative for sharp enough barriers---there is enough to be gained by getting off the top of the barrier that the field can profitably roll off to different sides of the barrier on different sides of the sphere. This contributes a total of four more negative modes. But these modes turn positive for barriers that are too flat: indeed they turn positive when $|V''|=4H_{\textrm{top}}^2$, exactly the moment when the CDLs merge with the HM. 

This story repeats for yet sharper barriers. As $|V''|$ is raised past $4 \ell H_{\textrm{top}}^2$, the $\ell$th spherical harmonic modes of the HM go negative, and, simultaneously, the oscillating bounce solutions of \cite{Hackworth:2004xb}  emerge from the  HM in a cusp catastrophe.\footnote{In addition to annihilating at the HM in a cusp, oscillating bounces can annihilate away from the HM in a fold \cite{Hackworth:2004xb,Batra:2006rz}.  In fact, in some parameter regimes, the CDL instanton itself never merges with the HM instanton; instead, the HM instanton sends out an oscillating bounce (at a cusp), which then annihilates the CDL instanton (at a fold).  } 

\section{Summary}

We cracked the case. Instantons can disappear smoothly, with their action becoming infinite, or abruptly, with their action staying finite.  We have seen subdominant instantons abruptly disappear when they are annihilated by other Euclidean solutions with higher action and one more negative mode.  The simplest and most generic abrupt disappearances are fold catastrophes, in which one minimum and one saddle annihilate.  But higher symmetry can give rise to higher-order catastrophes, and we saw that annihilations involving Hawking-Moss solutions are cusps. 
The 6D Einstein-Maxwell theory discussed in the Appendix exhibits a  complicated pattern of appearing and disappearing instantons.

An abruptly disappearing instanton is never the fastest decay out of the $A$ vacuum, but it can be the fastest, or indeed the only, decay into the $C$ vacuum.  Not only can a small change in the parameters of the theory cause a precipitous decline in the tunneling rate to $C$, but  this change can be \emph{lowering}  the potential barrier.  Let's say it the other way around: raising the barrier can make tunneling to $C$ faster.

\section*{Acknowledgements}

Thanks to Matt Johnson, Paul Steinhardt, and particularly to Erick Weinberg.

\bibliographystyle{utphys}
\bibliography{mybib.bib}

\appendix

\section{6D Einstein-Maxwell Theory}

In this appendix, we see instanton disappearances at work in a much richer landscape. We consider the specific example of flux tunneling between the four-dimensional vacua of 6D Einstein-Maxwell theory, which exhibits both smooth and abrupt disappearances.  In \cite{Brown:2010mf}, we constructed thick-walled instanton profiles for this theory.  Here we will show that these disappearances operate by the same physics as our other examples.   This appendix is recommended  only for readers  familiar with \cite{Brown:2010mf}. 

The 6D Einstein-Maxwell theory gives rise to a landscape of four-dimensional vacua, indexed by the flux $F^2$ wrapping the internal two-sphere.  The flux  contributes to the effective potential for the radion $\psi$.  For large $F^2$, the four-dimensional vacuum is de Sitter and the radion is unstable to tunneling out towards $\psi\rightarrow\infty$, which corresponds to decompactification.  For smaller $F^2$, the four-dimensional vacuum is Minkowski or AdS and the vacuum is stable against decompactification. Transitions between the compactified vacua proceed by flux tunneling, nucleating a bubble of charged brane with tension $T$ that changes the flux in the interior by an amount $\Delta F^2$.   The change in the flux decreases the energy density in the interior, and shifts the value of the minimum of the radion. The tension of the brane makes a finite spike in the radion profile.

Let's consider tunneling out of four-dimensional vacua with various values of $F^2$. We treat three cases in order of increasing complexity,  characterized by the number of ways to decompactify without changing the flux. Case 1 is decay from Minkowski or AdS, which cannot decompactify. 
Case 2 is decay from high de Sitter, which can decompactify by a Hawking-Moss (HM) instanton, for which $\psi$ sits at the crest of the potential for the full horizon volume. Case 3 is decay from low de Sitter, which can decay by Coleman-De Luccia bubble nucleation; in this case there is also a HM solution, though as we have seen it has extra negative modes. The two de Sitter cases are divided by the critical value of $V$ at which the CDL and HM instantons merge. 
(We ignore oscillating bounces \cite{Hackworth:2004xb}, which are known to be universally subdominant.)  

Our approach here will be orthogonal to that of \cite{Brown:2010mf}.  There, we held fixed the relationship between $T$ and $\Delta F^2$, and varied the true and false vacuum energies. Here, we will hold fixed the true and false vacuum energies, and vary $T$.

\subsection*{Case 1: Decays from Minkowski and AdS}

\begin{figure}[htbp] %  figure placement: here, top, bottom, or page
    \centering
    \includegraphics[width=4in]{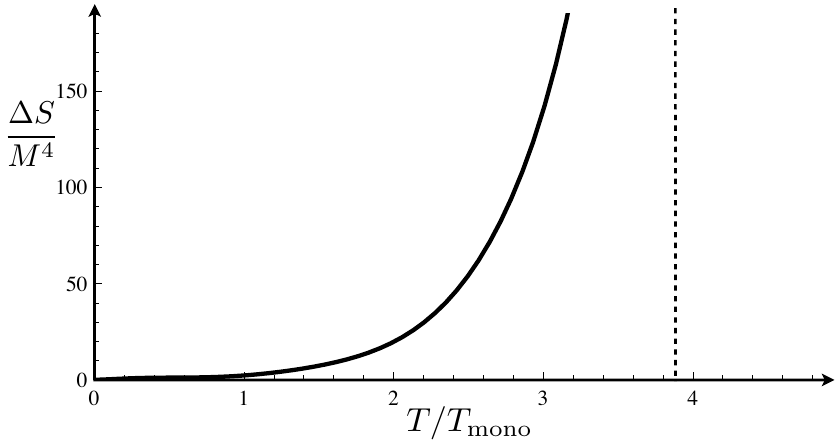} 
    \caption{The tunneling exponent as a function of the tension $T$ of the interpolating brane for decay from Minkowski space to a fixed AdS true vacuum.  $T_\text{mono}$ corresponds to the tension of the so-called monoflux brane of \cite{Brown:2010mf}, but here it just provides a reference tension. $M$ is the 4D Planck mass in units of the 6D Planck mass. There is a critical value of $T$, indicated by the dotted line, at which $\Delta S$ diverges and the instanton smoothly disappears by gravitational blocking.}
    \label{BofTMink}
 \end{figure}

\begin{figure}[htbp] %  figure placement: here, top, bottom, or page
    \centering
    \includegraphics[width=4in]{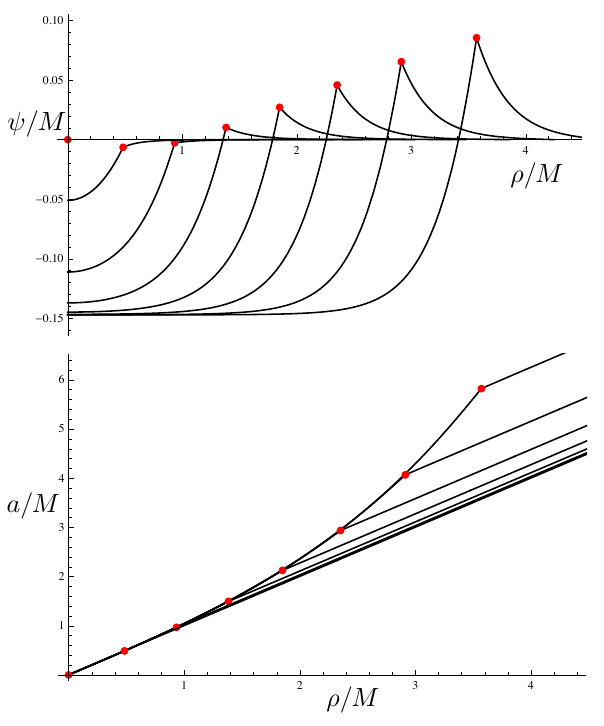} 
    \caption{Shown superimposed are the instanton profiles that mediate decay from Minkowski space to a fixed AdS true vacuum, as the tension of the interpolating brane is varied.  As the tension increases, the brane, indicated by a red dot, moves to larger $\rho$ and the spike in the vicinity of the brane grows.  The next profile in the plot would have greater than the critical value of $T$, at which both $a(\bar{\rho})$ and $\bar{\rho}$ diverge.}
    \label{varyTMink}
 \end{figure}

Figures~\ref{BofTMink} and \ref{varyTMink} show the rate and instanton profiles for decays from Minkowski to a fixed AdS true vacuum, for different values of $T$.  (Similar plots can be drawn for AdS false vacua.)
When $T=0$ there is no barrier to bubble nucleation: the brane sits at the origin $\bar{\rho} =a(\bar\rho)= 0$, $\psi$ sits everywhere in the false vacuum value ($\psi = 0$) and $\Delta S=0$. As $T$ increases, the radius of the bubble, the spike in the vicinity of the brane, and $\Delta S$ all grow.  At a critical value of $T$, $a(\bar{\rho})$ and $\Delta S$ diverge. For larger $T$, the instanton has smoothly disappeared.  This is the gravitational blocking we described in Sec.~\ref{sec:gravitationalblocking}. 

%For every true vacuum, there is a corresponding critical value of  $T$. Monoflux branes are always lighter than this value, so no monoflux transitions are gravitationally blocked. Multiflux branes are always heavier than monoflux branes, but for giant leaps, they too are lighter than this critical value.  But for small steps, multiflux branes are heavier than this value and gravitationally blocked: as you dial $T$ from the monoflux towards the multiflux values, the rate goes smoothly to zero and the instanton disappears. 
 
\subsection*{Case 2: Decays from high de Sitter}

In de Sitter space, energy is not conserved and no decay is gravitationally blocked.  There are no smooth disappearances, but there are still abrupt disappearances. 

\begin{figure}[htbp] %  figure placement: here, top, bottom, or page
    \centering
    \includegraphics[width=6.5in]{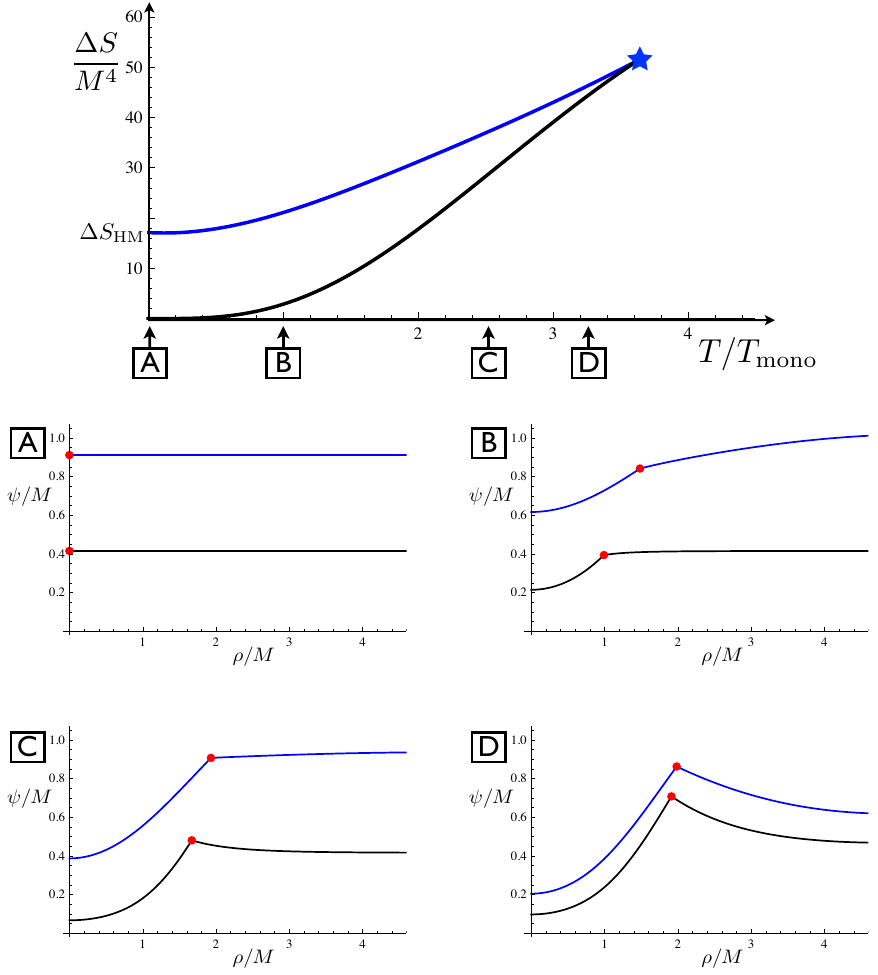} 
    \caption{Case 2, decay from high de Sitter. Top pane: the tunneling exponent $\Delta S$ as a function of the brane tension $T$ for decay from a fixed high de Sitter to a fixed true vacuum.  Lower panes: sample instanton profiles are shown for four different values of $T$.  The red dot indicates the location of the charged brane. The solution with lower action and smaller $\psi$ is drawn in black; it is the flux tunneling instanton. The other solution, with higher action and larger $\psi$, is drawn in blue; this solution has an extra negative mode.  As $T$ is increased, the solutions approach, merge, and annihilate at the star.  For higher $T$, the instanton has abruptly disappeared.}
    \label{HighdSDisappear}
 \end{figure}

Figure~\ref{HighdSDisappear} shows the rate and instanton profiles for decays from high de Sitter to a fixed true vacuum, for different values of $T$. When $T=0$, the brane again sits at the origin $\bar{\rho}=a(\bar\rho)=0$, but $\psi$ now has two options. It can sit, as before, in the false vacuum: this has $\Delta S=0$ and a single negative mode that corresponds to the growth of the bubble. This solution is drawn in black and mediates flux tunneling. Or it can sit at the crest of barrier, as in the HM solution: this has $\Delta S=\Delta S_{\text{HM}}$ and two negative modes corresponding to the bubble growing and to $\psi$ falling off the crest of the hill and rolling in one direction to decompactification and in the other into the black solution.  This solution is drawn in blue and is not an instanton because of the extra negative mode.

As $T$ increases, the two solutions approach one another.  The black solution  grows more spiked, and the blue solution moves to lower $\psi$, because it can rely on the heavy brane to help it decompactify.  Eventually, at a critical value of $T$, the two solutions merge and annihilate.
For this example, when the flux tunneling instanton is annihilated it is subdominant to pure HM decompactification: the blue solution has $\Delta S=\Delta S_{\text{HM}}$ at $T=0$, and then grows with $T$, so the star must lie above $\Delta S_\text{HM}$.

 \subsection*{Case 3: Decays from low de Sitter}
 
   \begin{figure}[htbp] %  figure placement: here, top, bottom, or page
    \centering
    \includegraphics[width=6.5in]{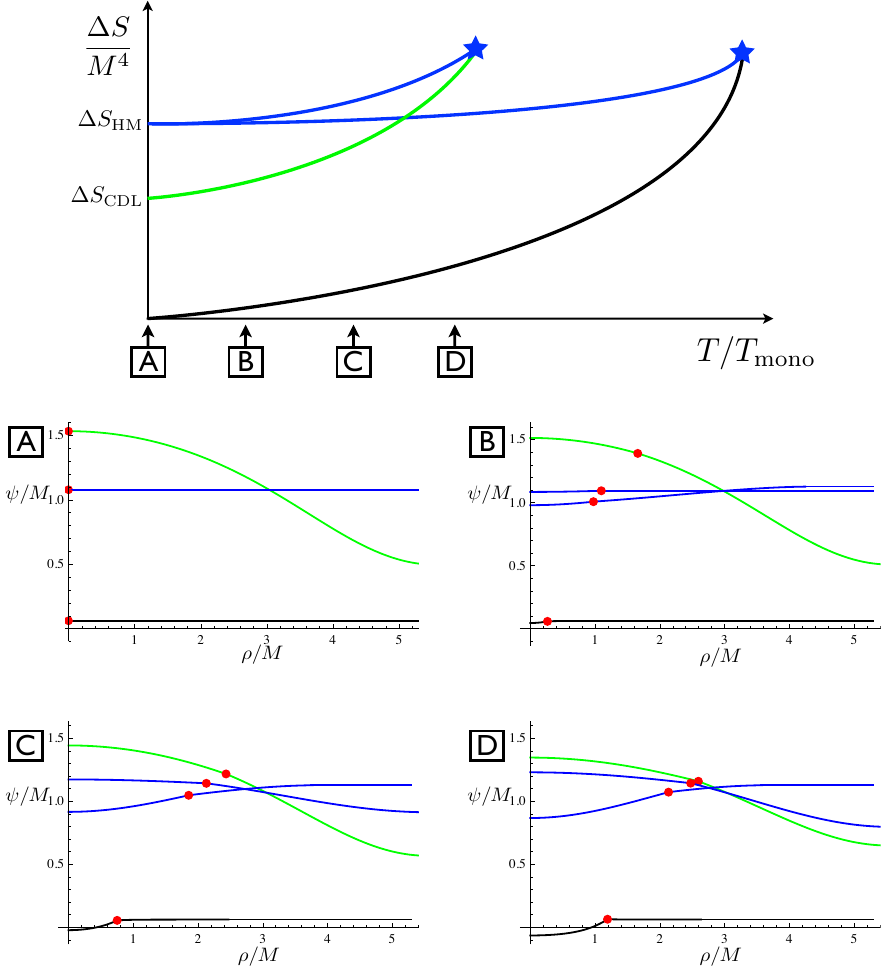} 
    \caption{Case 3, decay from low de Sitter. Top pane: a cartoon of the tunneling rate $\Delta S$ as a function of the brane tension $T$ for decay from a fixed low de Sitter to a fixed true vacuum. (On the non-cartoon plot, it is hard to see both branches at once.)  Lower panes: sample instanton profiles are shown for four different values of $T$. The red dot indicates the location of the charged brane. There are four solutions.  The black one (low $\Delta S$, low $\psi$) is the standard flux tunneling instanton; the green one (intermediate $\Delta S$, high $\psi$) is the instanton that mediates simultaneous decompactification and flux tunneling.  The two blue solutions (high $\Delta S$, intermediate $\psi$) have extra negative modes; one solution annihilates the green instanton, the other solution eventually annihilates the black instanton.}
    \label{LowdSDisappear}
\end{figure}
 
Figure~\ref{LowdSDisappear} shows the rate and instanton profiles for decays from low de Sitter to a fixed true vacuum, for different values of $T$. When $T=0$, the brane again sits at the origin $a(\bar{\rho}) = \bar{\rho}=0$, but $\psi$ now has three choices. It can sit, as before, in the false vacuum: this has $\Delta S=0$ and a single negative mode that corresponds to the growth of the bubble. This solution is again drawn in black and mediates flux tunneling. Or it can adopt the CDL decompactification profile: this has $\Delta S=\Delta S_\text{CDL}$ and also one negative mode that corresponds to simultaneous flux tunneling and decompactification. This solution is drawn in green. Finally, it can sit at the crest of barrier, as in the HM solution: this has $\Delta S= \Delta S_{\text{HM}}$ and has a great many negative modes corresponding to the growth of the bubble, rolling away to decompactification and falling into the CDL solution. This solution is drawn in blue. 

There are now two bona fide tunneling instantons that will disappear: the black and the green. As $T$ increases, the blue solution splits immediately into two, one of which will move up and annihilate the green solution, the other of which will move down and eventually annihilate the black solution. 
We can see that simultaneous decompactification and flux tunneling is always subdominant to straightforward decompactification: the rates agree at $T=0$, but $\Delta S$ must increase with $T$.

\end{document}